\pdfoutput=1

\documentclass[11pt]{article}

\usepackage[final]{acl}

\usepackage{times}
\usepackage{latexsym}

\usepackage[T1]{fontenc}

\usepackage[utf8]{inputenc}

\usepackage{microtype}

\usepackage{inconsolata}

\usepackage{graphicx}

\usepackage{tabularx}
\usepackage{booktabs}
\usepackage{algorithm}
\usepackage{algorithmic}
\usepackage{multirow}
\usepackage{url}

\title{Cross-Domain Audio Deepfake Detection: Dataset and Analysis}

\author{Yuang Li, Min Zhang, Mengxin Ren, Miaomiao Ma, Daimeng Wei, Hao Yang \\
        Huawei Translation Services Center, China \\ \{liyuang3,  zhangmin186, renmengxin2, mamiaomiao, \\ weidaimeng, yanghao30\}@huawei.com}

\begin{document}

\maketitle

\begin{abstract}
Audio deepfake detection (ADD) is essential for preventing the misuse of synthetic voices that may infringe on personal rights and privacy. Recent zero-shot text-to-speech (TTS) models pose higher risks as they can clone voices with a single utterance. However, the existing ADD datasets are outdated, leading to suboptimal generalization of detection models. In this paper, we construct a new cross-domain ADD dataset comprising over 300 hours of speech data that is generated by five advanced zero-shot TTS models. To simulate real-world scenarios, we employ diverse attack methods and audio prompts from different datasets. Experiments show that, through novel attack-augmented training, the Wav2Vec2-large and Whisper-medium models achieve equal error rates of 4.1\% and 6.5\% respectively. Additionally, we demonstrate our models' outstanding few-shot ADD ability by fine-tuning with just one minute of target-domain data. Nonetheless, neural codec compressors greatly affect the detection accuracy, necessitating further research. Our dataset is publicly available~\footnote{\url{https://github.com/leolya/CD-ADD}}.
\end{abstract}

\section{Introduction}

Audio deepfakes, created by text-to-speech (TTS) and voice conversion (VC) models, pose severe risks to social stability by spreading misinformation, violating privacy, and undermining trust. For advanced TTS models, the subjective score of the synthetic speech can surpass that of the authentic speech~\cite{ju2024naturalspeech} and humans are often unable to recognize deepfake audio~\cite{muller2022does, cooke2024good}. Consequently, it is imperative to develop robust audio deepfake detection (ADD) models capable of identifying imperceptible anomalies.

Several datasets built upon various TTS and VC models have been released to benchmark the ADD task~\cite{yi2022add, yamagishi2021asvspoof, frank2021wavefake, wang2020asvspoof, yi2023add}. However, these datasets mainly include the traditional TTS models rather than the emerging zero-shot TTS models. Moreover, there is a lack of transparency regarding the specific types of models used within these datasets, hindering comprehensive analysis of cross-model performance. Additionally, the range of attacks these datasets consider is confined to conventional methods, excluding attacks associated with deep neural networks (DNNs), such as noise reduction and neural codec models. Based on the aforementioned datasets, a multitude of detection models have been proposed. These models incorporate diverse features, such as the traditional linear frequency cepstral coefficient~\cite{yan2022audio} and features derived from self-supervised learning~\cite{zeng2023deepfake, martin2022vicomtech}, emotion recognition~\cite{conti2022deepfake}, and speaker identification models~\cite{pan2022speaker}. These studies mainly concentrate on a single benchmark dataset. To demonstrate generalization capabilities, several studies have implemented cross-dataset evaluation~\cite{muller2022does, ba2023transferring}. Furthermore, to enhance the models' generalizability, researchers have explored the combination of data from various sources~\cite{kawa2022attack} and the integration of multiple features~\cite{yang2024robust}.

In this paper, we present a novel cross-domain ADD (CD-ADD) dataset, which encompasses more than 300 hours of speech data generated by five cutting-edge, zero-shot TTS models. We test nine different attacks, including those involving DNN-based codecs and noise reduction models. For cross-domain evaluation, rather than adopting the naive cross-dataset scenario, we formulate a unique task for zero-shot TTS models by analyzing pairwise cross-model performance and utilizing audio prompts from different domains. Experiments reveal: 1) The cross-domain task is challenging. 2) Training with attacks improves adaptability. 3) The ADD model is superior in the few-shot scenario. 4) The neural codec poses a major threat.
\begin{figure}[t]
  \centering
  \includegraphics[width=\linewidth]{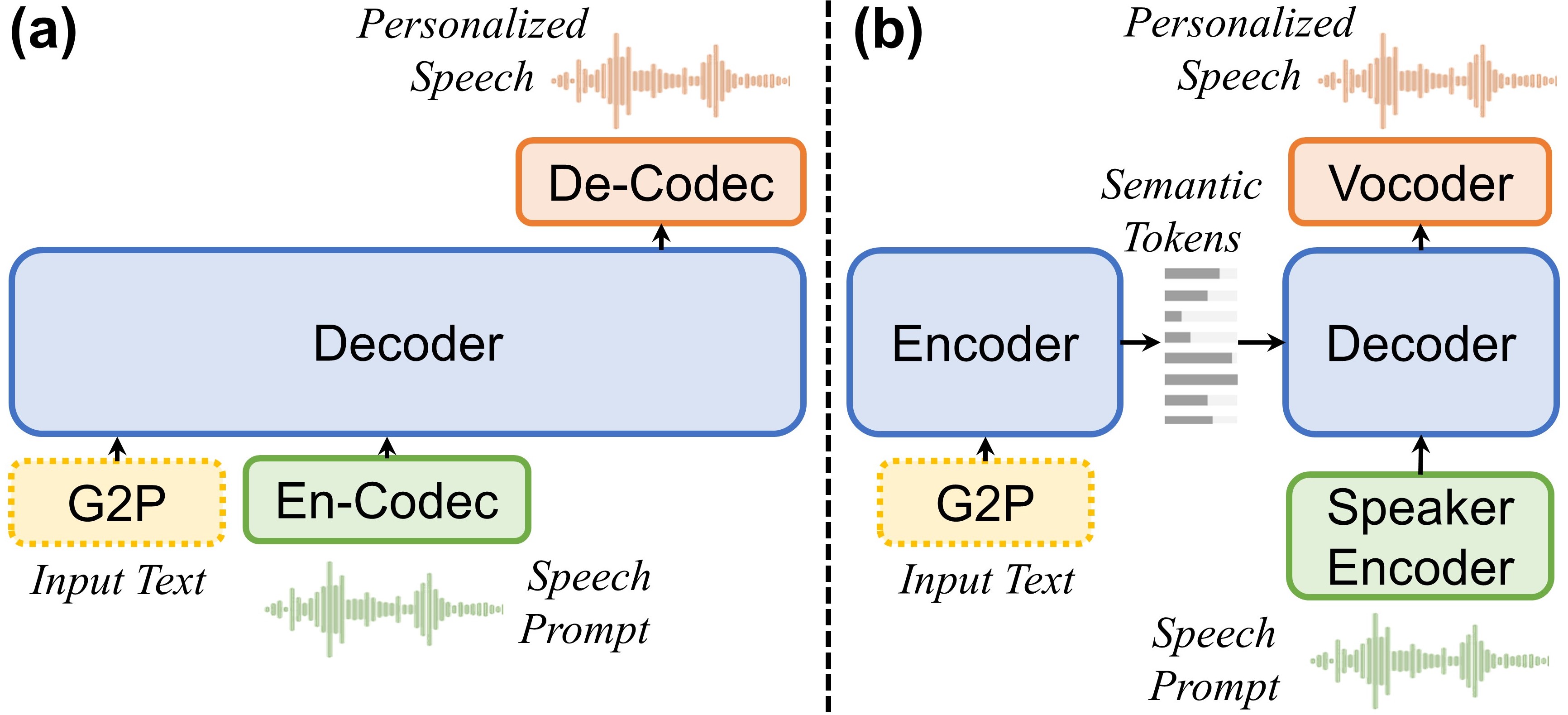}
  \caption{Zero-shot TTS architectures. a) Decoder-only. b) Encoder-decoder.}
  \label{fig:arch}
\end{figure}

\section{Methods}

\subsection{Dataset Construction}
\label{sec:data}

As shown in Figure~\ref{fig:arch}, we can categorize the zero-shot TTS models into two types: \textbf{1) Decoder-only} (VALL-E~\cite{wang2023neural}): It accepts phoneme representations and the speech prompt's discrete codes as input, and generates output speech codes autoregressively. These codes are transformed into personalized speech signals. \textbf{2) Encoder-decoder} (YourTTS~\cite{casanova2022yourtts}, WhisperSpeech~\cite{kharitonov2023speak}, Seamless Expressive~\cite{barrault2023seamless}, and OpenVoice~\cite{qin2023openvoice}): An encoder extracts semantic information, while a decoder incorporates speaker embeddings from the speech prompt. Together with the vocoder, the autoregressive (AR) or non-autoregressive (NAR) decoder generates personalized speeches. When the encoder is trained to remove speaker-specific information from the input speech, it transforms into a VC model.

For zero-shot TTS, AR decoding may introduce instability, leading to errors such as missing words. Poor-quality speech prompts, characterized by high noise levels, can result in unintelligible output. To address this, we enforce quality control during dataset construction (Algorithm~\ref{alg:dataset}). Specifically, we utilize an automatic speech recognition (ASR) model to predict the transcription of the generated speech. If the character error rate (CER) exceeds the threshold, we regenerate the speech using alternative prompts. Utterances are discarded if the CER remains above the threshold after a predefined number of retries. Prompts from different domains are used to evaluate the generalizability of ADD models. Our dataset introduces two tasks: \textbf{1) In-model ADD} considers all models during both training and testing. \textbf{2) Cross-model ADD} excludes data from one TTS model during training and uses data from this TTS model only during testing.

ADD models should generalize to in-the-wild synthetic data, which requires a well-designed cross-model evaluation that can represent the real-world scenario. To select the appropriate TTS model for testing, we conduct a pairwise cross-model evaluation, where the Wav2Vec2-base model is trained exclusively on the data produced by a single TTS model and subsequently evaluated on the datasets generated by alternative TTS models. We identify the TTS model that poses the greatest challenge, as evidenced by the high equal error rate (EER), and use it as the test set.

\begin{algorithm}[t]
\caption{Dataset construction}
\small
\begin{algorithmic}[1]
\REQUIRE $ prompts, text, retry, threshold $
\STATE $i \gets 0$
\STATE $success \gets False$
\WHILE{$i < retry$}
    \STATE $p \gets random\_select(prompts)$
    \STATE $audio \gets TTS(text, p)$
    \STATE $\hat{text} \gets ASR(audio)$
    
    \IF{$CER(text, \hat{text}) < threshold$}
        \STATE $success \gets True$
        \STATE \textbf{break}
    \ENDIF
    \STATE $i \gets i + 1$
\ENDWHILE
\RETURN $audio, success$
\end{algorithmic}
\label{alg:dataset}
\end{algorithm}

\subsection{Attacks}

\begin{figure}[h]
  \centering
  \includegraphics[width=0.9\linewidth]{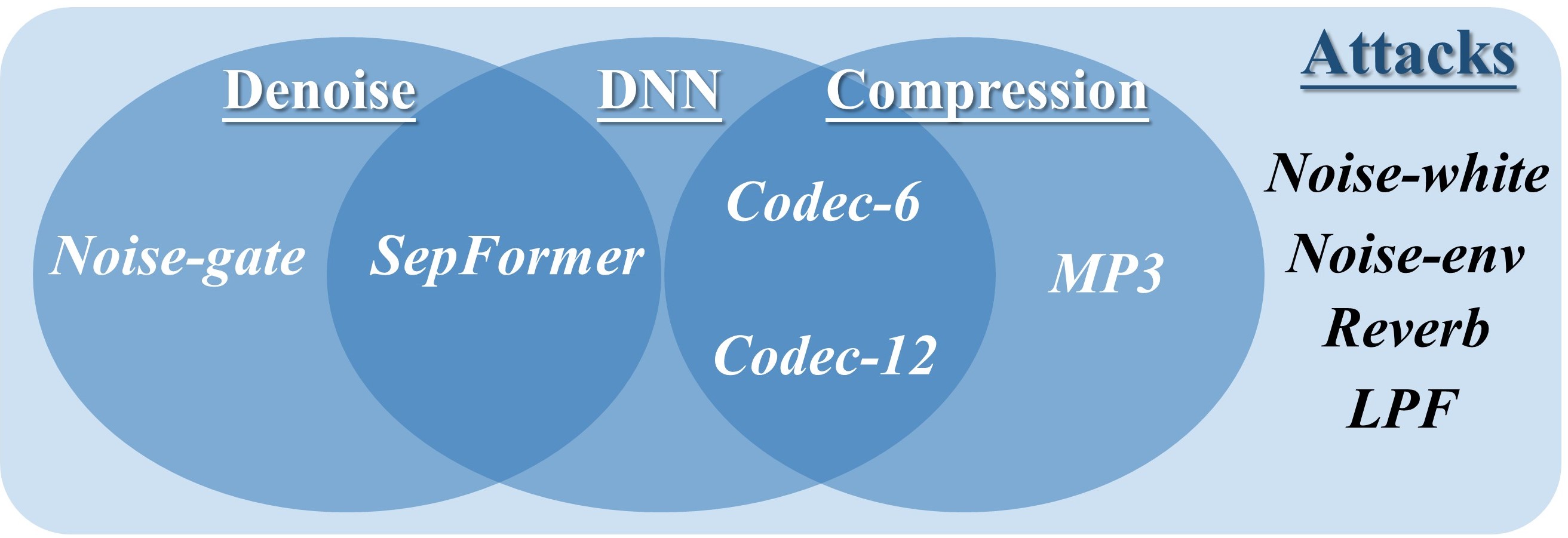}
  \caption{Categories of tested attacks.}
  \label{fig:attack}
\end{figure}


Figure~\ref{fig:attack} presents the nine attacks we test. For traditional attacks, we add white Gaussian noise (Noise-white) and environmental noise (Noise-env)~\cite{maciejewski2020whamr} with a signal-to-noise ratio ranging from 15dB to 20dB, use artificial reverberation (Reverb) with a duration of 0.2 to 0.4 seconds, and apply a low-pass filter (LPF) within the 4kHz to 8kHz range. Furthermore, we employ lossy compression methods such as MP3 and a DNN-based Encodec model~\cite{defossez2022high} operating at bit rates of 6kbps (Codec-6) and 12kbps (Codec-12). In terms of noise reduction, we utilize the conventional noise gate approach to eliminate stationary noise and the time-domain SepFormer model~\cite{subakan2021attention}.

\subsection{ADD Methods}

We fine-tune pre-trained speech encoders for the ADD task, namely,  Wav2Vec2~\cite{baevski2020wav2vec} and the Whisper encoder~\cite{radford2022robust}. We merge multi-layer features by using learnable weights, and employ a classifier head with two projection layers and one global pooling layer to obtain the final logits. To adapt the model to attacks, we consider all attacks with the same probability on-the-fly during training. We also consider a few-shot scenario, where we extend the cross-model evaluation by fine-tuning the ADD model with just one minute of target-domain speech data. This experiment simulates a situation where only the limited synthetic speech from a TTS model is available, such as the speech from a demo website or a single video.

\section{Experimental Setups}

The training set for the CD-ADD dataset was generated using the train-clean-100 subset of LibriTTS~\cite{zen2019libritts}, and the dev-clean and test-clean subsets of LibriTTS, along with the test set of TEDLium3~\cite{hernandez2018ted}, were utilized for the evaluation datasets. The transcriptions were used as the input text, and the real speech signals were used as the real samples and the speech prompts. For dataset construction, we used the five zero-shot TTS models mentioned in Section~\ref{sec:data}, a CER threshold of 10\%, and a maximum retry limit of five. For cross-model evaluation, the speech from Seamless Expressive served as the test set. Appendix~\ref{apx:A} provides comprehensive details on the TTS model checkpoints and the models used for attacks, and Appendix~\ref{apx:B} presents the specific statistics of the CD-ADD dataset that is comprised of over 300 hours of training data and 50 hours of test data.

For the ADD task, we combined our CD-ADD dataset with the ASVSpoof2019~\cite{wang2020asvspoof} training set and fine-tuned the base model, which includes Wav2Vec2~\cite{baevski2020wav2vec} and the Whisper encoder~\cite{radford2022robust}, for four epochs with a learning rate of $3e-5$ and a batch size of 128. For attack-augmented training, we increased the number of epochs to eight, as the model converges more slowly due to attacks. The probability of each attack was 10\% and only one attack type was used for each utterance. For the evaluation metric, we adopted the widely used equal error rate (EER).
\section{Experimental Results}

\subsection{Pairwise Cross-Model Evaluation}

\begin{figure}[t]
  \centering
  \includegraphics[width=\linewidth]{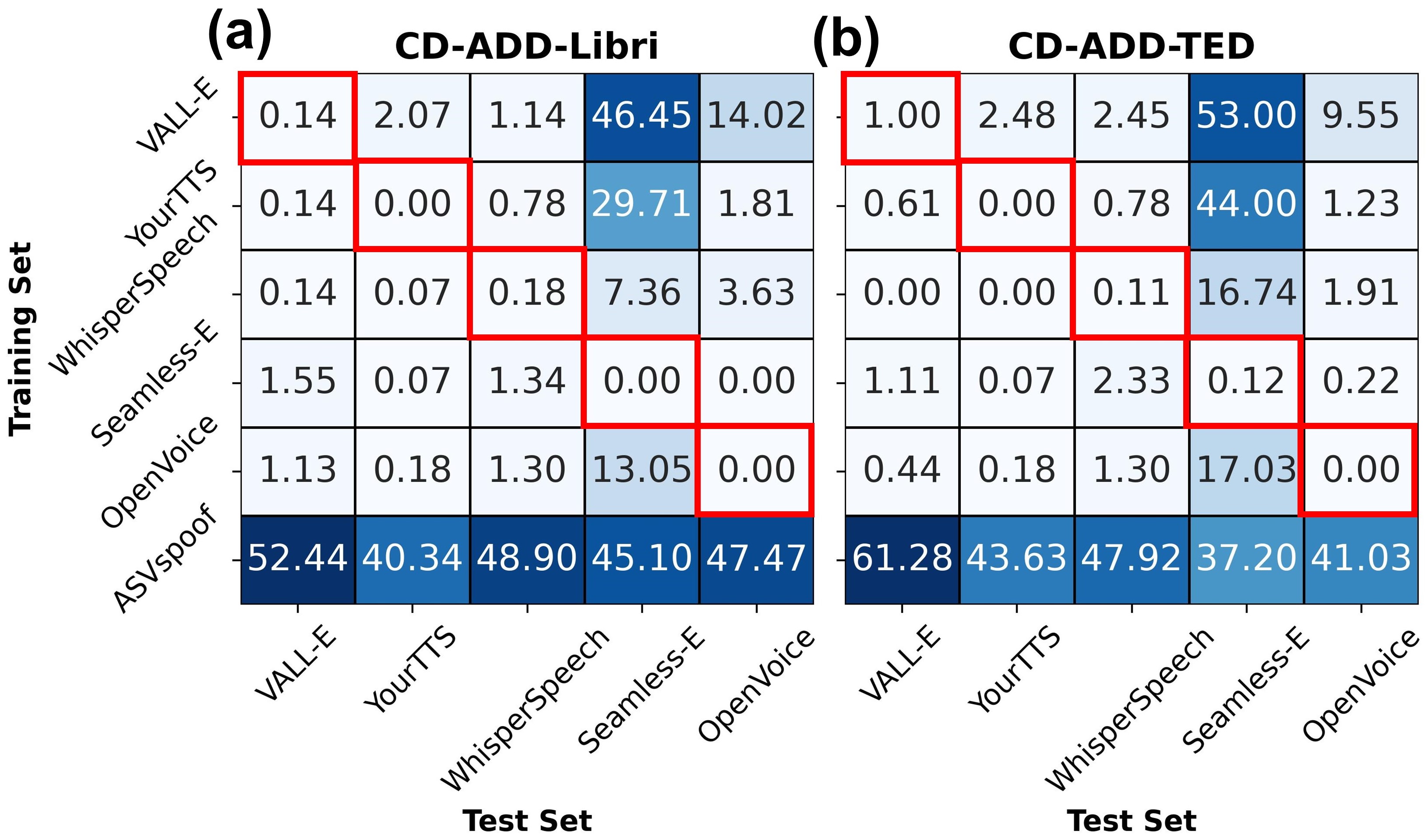}
  \caption{Cross-model EER matrix, where the Wav2Vec2-base model was trained using data generated from a single TTS model and subsequently evaluated on data originating from other TTS models.}
  \label{fig:matrix}
\end{figure}

\begin{table}[t]
\centering
\scalebox{0.75}{
\begin{tabular}{c | c | c c } 
\toprule
 & Training data & Libri & TED \\
 \midrule
 In-model & CD-ADD & 0.11 & 0.35 \\
 & CD-ADD + ASVspoof & \textbf{0.07} & \textbf{0.12} \\
 \midrule
 Cross-model & CD-ADD & 12.14 & \textbf{20.34} \\
 & CD-ADD + ASVspoof & \textbf{7.85} & 21.40 \\
\bottomrule
\end{tabular}
}
\caption{Performance of Wav2Vec2-base measured by EER (\%).}
\label{tab:result1}
\end{table}

As illustrated in Figure~\ref{fig:matrix}, the pairwise evaluation indicates that the ADD system exhibits optimal performance when both the training and testing sets are derived from the same TTS model. This trend holds true irrespective of the speech prompts' domain (whether they originate from the in-domain LibriTTS dataset or the cross-domain TEDLium dataset), with the EERs consistently remaining below 1\%. However, in the cross-model evaluation, the EERs vary significantly among different TTS model combinations. For example, the Wav2Vec2-base model fine-tuned with YourTTS-synthesized data can generalize to VALL-E-synthesized data, achieving EERs of 0.14\% and 0.61\% for the Libri and TED subsets of the CD-ADD test sets, respectively. However, it struggles to generalize to the Seamless Expressive model, resulting in much higher EERs of 29.71\% and 44.00\%. This indicates that randomly choosing a test set whose speech data is generated by a TTS model could result in overestimated generalizability of the ADD model, due to shared artifacts between TTS models and potential overfitting. Therefore, we selected Seamless Expressive as the test set as it has notably high EERs. It is worth noting that the model trained on the prevalent ASVSpoof dataset fails to generalize to the zero-shot TTS models.  However, combining ASVspoof with the CD-ADD dataset can slightly improve the performance (Table~\ref{tab:result1}), so these two datasets are combined by default. 


\subsection{Comparisons Between Attacks}

\begin{table}[t]
\centering
\scalebox{0.75}{
\begin{tabular}{c | c c | c c} 
\toprule
\multirow{2}*{Attack} & \multicolumn{2}{c|}{In-model} & \multicolumn{2}{c}{Cross-model} \\ 
 &  & + Aug. &  & + Aug. \\
 \midrule
Baseline & 0.1 / 0.1 & 0.0 / 0.1 & 7.9 / 21.4 & 5.0 / 10.1 \\
Noise-white & 9.4 / 9.1 & 0.8 / 0.7 & 34.7 / 45.0 & 9.9 / 10.3 \\
Noise-env & 9.0 / 4.7 & 0.5 / 0.3 & 29.2 / 31.1 & 9.4 / 9.3 \\
Reverb & 13.0 / 17.1 & 1.1 / 1.2 & 29.6 / 33.1 & 18.1 / 23.7 \\
LPF & 1.3 / 1.2 & 0.1 / 0.3 & 14.3 / 23.4 & 6.6 / 8.9 \\
MP3 & 0.3 / 0.2 & 0.0 / 0.1 & 13.2 / 22.1 & 5.4 / 8.3 \\
Codec-12 & 2.9 / 1.4 & 0.3 / 0.3 & 21.4 / 31.0 & 11.4 / 18.3 \\
Codec-6 & 7.4 / 5.2 & 0.9 / 1.2 & 30.5 / 35.2 & \underline{18.5} / \underline{28.9} \\
Noise-gate & 11.8 / 6.5 & 0.9 / 1.1 & 33.7 / 27.7 & 12.3 / 14.5 \\
SepFormer & 1.0 / 2.8 & 0.1 / 0.4 & 9.2 / 12.6 & \textbf{3.3} / \textbf{5.5} \\
\bottomrule
\end{tabular}
}
\caption{Performance of Wav2Vec2-base under various attacks measured by EER (\%) on Libri and TED test sets respectively. "+Aug." indicates all attacks are included during training.}
\label{tab:result2}
\end{table}

\begin{figure}[t]
  \centering
  \includegraphics[width=\linewidth]{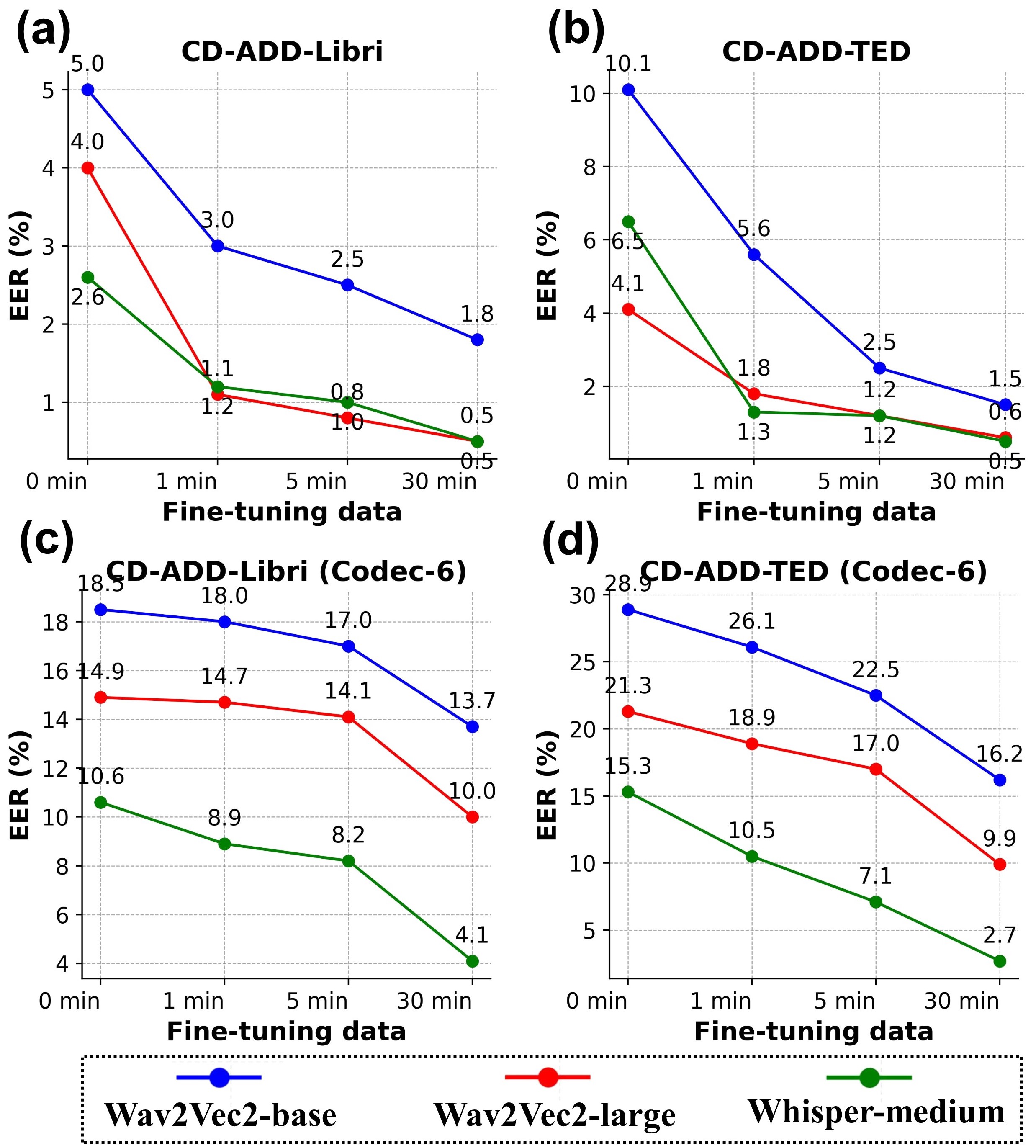}
  \caption{Few-shot performance of three base models measured by EER (\%).}
  \label{fig:fewshot}
\end{figure}

As shown in Table~\ref{tab:result2}, without augmentation, all attacks negatively impact the model, with more noticeable effects in cross-model configurations. With attack-augmented training, the Wav2Vec2-base model demonstrates resilience against most attacks. In the in-model setup, the EERs of the attacked models are only slightly higher than the baseline. In the cross-model setup, a significant decrease in EERs is observed for the augmented model compared to the non-augmented model. Notably, certain attacks improve the ADD model's generalizability, as indicated by the reduced EERs in the TED subset. For example, compared with the EER of 10.1\% for the baseline, the LPF reduces the EER to 8.9\%, the MP3 compression reduces the EER to 8.3\%, and the SepFormer reduces the EER to 5.5\%. All these attacks remove spectral information and force the ADD model to rely more on features from the low-frequency band, thus mitigating overfitting. However, certain attacks, such as reverberation and the Encodec, lead to relatively high EERs. The encoder-decoder architecture and the vector quantization of the Encodec, especially at lower bit rates, have the potential to obliterate essential features for detecting synthetic speeches.

\subsection{Results of Few-Shot Fine-Tuning}

Figure~\ref{fig:fewshot} compares the cross-model ADD performance of three base models: Wav2Vec2-base, Wav2Vec2-large, and Whisper-medium. The Wav2Vec2-large and the Whisper-medium models have similar performance, notably superior to the Wav2Vec2-base model (Figure~\ref{fig:fewshot} (a, b)). With the most challenging Encodec attack, the Whisper model performs significantly better than the Wav2Vec2 models (Figure~\ref{fig:fewshot} (c, d)). We can also observe that with only one minute of in-domain data from Seamless Expressive, the EER can be reduced significantly. This suggests that our models are capable of fast adaptation to in-the-wild TTS systems with just a few samples from a demo website or a video, which is crucial for real-world deployment. However, we find that in-domain fine-tuning is less effective when the audio is compressed with the Encodec, as the reduction in EER is less significant.

\section{Conclusion}

In conclusion, our study presents a CD-ADD dataset, addressing the urgent need for up-to-date resources to combat the evolving risks of zero-shot TTS technologies. Our dataset, comprising over 300 hours of data from advanced TTS models, enhances model generalization and reflects real-world conditions. This paper highlights the risks of attacks and the potential of few-shot learning in ADD, facilitating future research.
\section{Limitation}

The current CD-ADD dataset is limited to five zero-shot TTS models. Future expansions are planned to include a broader range of zero-shot TTS models, as well as conventional TTS and VC models, to improve the dataset diversity. Additionally, the attack-augmented training is constrained to a single attack per sample, with separate analysis conducted for each attack. Subsequent research will focus on investigating the effects of combined attacks. Furthermore, the performance in ADD tasks with audio compressed by neural codecs is suboptimal, requiring the development of optimization strategies and the exploration of more neural codec models.

\bibliography{custom}

\appendix

\section{Appendix: Open Source Tools}
\label{apx:A}
\textbf{Zero-shot TTS models:}

\begin{itemize}
    \item VALL-E: \url{https://github.com/Plachtaa/VALL-E-X}
    \item YourTTS: \url{https://github.com/coqui-ai/TTS}
    \item Seamless Expressive: \url{https://github.com/facebookresearch/seamless_communication}
    \item WhisperSpeech: \url{https://github.com/collabora/WhisperSpeech?tab=readme-ov-file}
    \item OpenVoice: \url{https://github.com/myshell-ai/OpenVoice}
\end{itemize}

\textbf{Base models:}

\begin{itemize}
    \item Wav2Vec2-base: \url{https://huggingface.co/facebook/wav2vec2-base}
    \item Wav2Vec2-large: \url{https://huggingface.co/facebook/wav2vec2-large}
    \item Whisper-medium: \url{https://huggingface.co/openai/whisper-medium}
\end{itemize}

\textbf{ASR model:}

\begin{itemize}
    \item HuBERT-large-CTC: \url{https://huggingface.co/facebook/hubert-large-ls960-ft}
\end{itemize}

\textbf{Attacks:}

\begin{itemize}
    \item Noise-gate: \url{https://github.com/timsainb/noisereduce }
    \item SepFormer: \url{https://huggingface.co/speechbrain/sepformer-whamr}
    \item Codec-6/12: \url{https://github.com/facebookresearch/encodec}
\end{itemize}

\section{Appendix: CD-ADD Dataset}
\label{apx:B}

Table~\ref{tab:data1} presents the statistics of the CD-ADD dataset. The average utterance length exceeds eight seconds, which is longer than that of traditional ASR datasets. The number of utterances for TTS models is less than that of real utterances because some synthetic utterances fail to meet the CER requirements. Among them, VALL-E has the fewest utterances due to the decoder-only model's relative instability. Table~\ref{tab:data2} compares five zero-shot TTS models in terms of the word-error-rate (WER) and speaker similarity. Speaker similarity is based on the LibriTTS test-clean subset, where ECAPA-TDNN is used to extract speaker embeddings. VALL-E and WhisperSpeech have the highest speaker similarity scores, while OpenVoice ranks lowest. Conversely, VALL-E achieves the highest WER, and OpenVoice has the lowest.

\begin{table*}[t]
\centering
\scalebox{0.85}{
\begin{tabular}{c | c c c | c c c | c c c | c c c} 
\toprule
& \multicolumn{3}{c|}{train-clean} & \multicolumn{3}{c|}{dev-clean} & \multicolumn{3}{c|}{test-clean} &  \multicolumn{3}{c}{test-TED}\\
& Num. & Total & Avg. & Num. & Total & Avg. & Num. & Total & Avg. & Num. & Total & Avg. \\
\midrule
Real & 18339 & 49.6 & 9.7 & 3111 & 8.2 & 9.5 & 2762 & 8.0 & 10.5 & 899 & 2.62 & 10.49 \\
\midrule
VALL-E & 15869 & 41.0 & 9.3 & 2770 & 7.1 & 9.2 & 2275 & 6.1 & 9.6 & 452 & 1.13 & 9.01\\
Seamless Expressive & 17829 & 42.6 & 8.6 & 3042 & 7.7 & 9.1 & 2717 & 8.0 & 10.6 & 816 & 2.11 & 9.32 \\
YourTTS & 18202 & 49.3 & 9.8 & 3093 & 8.2 & 9.5 & 2739 & 7.9 & 10.4 & 868 & 2.14 & 8.86\\
WhisperSpeech & 18300 & 54.8 & 10.8 & 3106 & 9.3 & 10.8 & 2760 & 8.9 & 11.6 & 862 & 2.71 & 11.33 \\
OpenVoice & 18024 & 40.9 & 8.2 & 3099 & 7.0 & 8.18 & 2753 & 6.7 & 8.8 & 883 & 1.99 & 8.13 \\
\bottomrule
\end{tabular}
}
\caption{The numbers of utterances (Num.), the total duration (Total), and the average duration of each utterance (Avg.) of the CD-ADD dataset.}
\label{tab:data1}
\end{table*}

\begin{table}[h]
\centering
\scalebox{0.95}{
\begin{tabular}{c | c c } 
\toprule
& WER $\downarrow$ & Spk. $\uparrow$ \\
\midrule
Real & 2.4 & 1.00 \\
\midrule
VALL-E & \underline{10.1} & \textbf{0.56} \\
Seamless Expressive & 5.3 & 0.52 \\
YourTTS & 5.4 & 0.53 \\
WhisperSpeech & 3.2 & \textbf{0.56} \\
OpenVoice & \textbf{2.6} & \underline{0.36} \\
\bottomrule
\end{tabular}
}
\caption{Zero-shot TTS performance measured by WER (\%) and speaker similarity (Spk.).}
\label{tab:data2}
\end{table}

\end{document}